\documentclass[epj]{svjour}
\usepackage{graphics}
\usepackage{epsfig}
\def\proof{\noindent{\sl Proof:}\kern0.6em}

\def\frac#1#2{\hbox{$#1\over#2$}}
\def\dual{\mathstrut^*\kern-0.1em}

\def\lvec#1{\setbox0=\hbox{$#1$}
    \setbox1=\hbox{$\scriptstyle\leftarrow$}
    #1\kern-\wd0\smash{
    \raise\ht0\hbox{$\raise1pt\hbox{$\scriptstyle\leftarrow$}$}}
    \kern-\wd1\kern\wd0}
\def\rvec#1{\setbox0=\hbox{$#1$}
    \setbox1=\hbox{$\scriptstyle\rightarrow$}
    #1\kern-\wd0\smash{
    \raise\ht0\hbox{$\raise1pt\hbox{$\scriptstyle\rightarrow$}$}}
    \kern-\wd1\kern\wd0}

\def\nabstar#1{\nabla\kern-0.5pt\smash{\raise 4.5pt\hbox{$\ast$}}
               \kern-4.5pt_{#1}}

\def\drvstar#1{\partial\kern-0.5pt\smash{\raise 4.5pt\hbox{$\ast$}}
               \kern-5.0pt_{#1}}

\def\MeV{{\rm MeV}}
\def\GeV{{\rm GeV}}

\def\fm{{\rm fm}}

\def\rhoprime{\rho\kern1pt'}
\def\rhobar{\bar{\rho}}
\def\rhobarprime{\rhobar\kern1pt'}
\def\rhobartilde{\kern2pt\tilde{\kern-2pt\rhobar}}
\def\rhobartildeprime{\kern2pt\tilde{\kern-2pt\rhobar}\kern1pt'}

\def\zetabar{\bar{\zeta}}
\def\zetaprime{\zeta\kern1pt'}
\def\zetabarprime{\zetabar\kern1pt'}

\def\diracstar#1#2{
    \setbox0=\hbox{$\gamma$}\setbox1=\hbox{$\gamma_{#1}$}
    \gamma_{#1}\kern-\wd1\kern\wd0
    \smash{\raise4.5pt\hbox{$\scriptstyle#2$}}}
\def\momp#1#2{
    \setbox0=\hbox{${#1}$}\setbox1=\hbox{${#1}_{#2}$}
    {#1}_{#2}\kern-\wd1\kern\wd0
    \smash{\raise4.5pt\hbox{$\scriptscriptstyle +$}}}
\def\momm#1#2{
    \setbox0=\hbox{${#1}$}\setbox1=\hbox{${#1}_{#2}$}
    {#1}_{#2}\kern-\wd1\kern\wd0
    \smash{\raise4.5pt\hbox{$\scriptscriptstyle -$}}}
\def\mompm#1#2{
    \setbox0=\hbox{${#1}$}\setbox1=\hbox{${#1}_{#2}$}
    {#1}_{#2}\kern-\wd1\kern\wd0
    \smash{\raise4.5pt\hbox{$\scriptscriptstyle \pm$}}}
\def\smomp#1#2{
    \setbox0=\hbox{${#1}$}\setbox1=\hbox{${#1}_{#2}$}
    {#1}_{#2}\kern-\wd1\kern\wd0
    \smash{\raise3pt\hbox{$\scriptscriptstyle +$}}}
\def\smomm#1#2{
    \setbox0=\hbox{${#1}$}\setbox1=\hbox{${#1}_{#2}$}
    {#1}_{#2}\kern-\wd1\kern\wd0
    \smash{\raise3pt\hbox{$\scriptscriptstyle -$}}}
\def\smompm#1#2{
    \setbox0=\hbox{${#1}$}\setbox1=\hbox{${#1}_{#2}$}
    {#1}_{#2}\kern-\wd1\kern\wd0
    \smash{\raise3pt\hbox{$\scriptscriptstyle \pm$}}}

\def\Vckm{V_{\rm CKM}}
\def\msbar{{\rm \overline{MS\kern-0.05em}\kern0.05em}}
\def\smallmsbar{\small\overline{\hbox{MS\kern-0.10em}}
                \hbox{\kern0.10em}}

\def\mpi{m_\pi}
\def\mK{m_{\rm K}}
\def\mps{m_{\rm PS}}

\def\mqu{m_{\rm u}}
\def\mqd{m_{\rm d}}
\def\mqs{m_{\rm s}}

\def\fK{f_{\rm K}}
\def\fB{f_{\rm B}}

\def\fBs{f_{\rm B_s}}
\def\fD{f_{\rm D}}

\def\fDs{f_{\rm D_s}}
\def\BB{B_{\rm B}}

\def\BBs{B_{\rm B_s}}
\def\BBhat{\widehat{B}_{\rm B}}

\def\BBshat{\widehat{B}_{\rm B_s}}

\newcommand{\be}{\begin{equation}}
\newcommand{\ee}{\end{equation}}
\newcommand{\bea}{\begin{eqnarray}}
\newcommand{\eea}{\end{eqnarray}}
\newcommand{\bi}{\begin{itemize}}
\newcommand{\ei}{\end{itemize}}

\newcommand{\lesssim}{\raisebox{-.6ex}{$\stackrel{\textstyle{<}}{\sim}$}}

\newcommand{\etal}{{\em et~al.}}

\newcommand{\nf}{N_{\rm f}}

\begin{document}
\title{Status of lattice calculations of $B$-meson decays and mixing}
\author{Hartmut Wittig
\thanks{Invited talk at Int. Europhysics Conf. on High Energy Physics
EPS-HEP2003, 17-23 July 2003, Aachen, Germany.}
}                     % Do not remove
\institute{DESY, Notkestra{\ss}e 85, D-22603 Hamburg, Germany}
\abstract{The present status of lattice calculations of $\fB$, $\BB$
and SU(3) flavour breaking ratios such as $\fBs/\fB$ is
reviewed. Particular attention is devoted to systematic uncertainties,
such as those arising from the lack of simulation data for dynamical
quarks with realistic masses, and the related difficulties associated
with chiral extrapolations. Global averages for decay constants and
mixing parameters are presented (Table\,\ref{tab_global}), and the
procedures to obtain them are discussed in detail.
\PACS{
      {11.30.Er}{CP violation} \and 
      {12.38.Gc}{Lattice QCD calculations}   \and
      {14.40.Nd}{Bottom mesons}
     } % end of PACS codes
}
\maketitle
\section{Introduction} \label{sec_intro}
%\vspace{-0.2cm}
A lot of activity is currently devoted to pin down the elements of the
CKM matrix $\Vckm$. In the Standard Model $\Vckm$ is unitary, which
implies triangle relations like
\be
 V_{\rm ud}V_{\rm ub}^*+V_{\rm cd}V_{\rm cb}^* +V_{\rm td}V_{\rm tb}^*=0.
\ee
Any deviation from unitarity is interpreted as a signature of ``new
physics''. In order to probe this scenario, experimental and
theoretical inputs are being used to over-constrain the elements of
$\Vckm$. However, relations between measurable quantities and CKM
matrix elements involving heavy quarks are usually afflicted with
large hadronic uncertainties. A typical example are the mass
differences $\Delta{M}_{\rm{d}}$ and $\Delta{M}_{\rm{s}}$ in the
$B^0\,\bar{B}^0$ and $B_{\rm{s}}^0\,\bar{B}^0_{\rm{s}}$ systems:
\bea
      & &
      \Delta M_{\rm{d}}={{G_{\rm F}^2 M_{\rm W}^2}\over{6\pi^2}}
      \eta_{\rm B}S(\frac{m_t}{M_{\rm W}})\,
      {\fB^2\BBhat}{\left|V_{\rm td}V_{\rm tb}^*\right|^2},
      \\
      & &
      {{\Delta M_{\rm{s}}}\over{\Delta M_{\rm{d}}}} = {\xi^2}
      {{m_{\rm{B_{s}}}}\over{m_{\rm{B}}}}
      {{|V_{\rm ts}|^2}\over{|V_{\rm td}|^2}},
      \quad \xi = {{\fBs\sqrt{\BBshat}}\over{\fB\sqrt{\BBhat}}}.
\eea
The limited accuracy with which the CKM elements on the rhs. are known
comes from theoretical uncertainties in the decay constants $\fB$,
$\fBs$ and the $B$-parameters $\BB$ and $\BBs$. These quantities have
been computed using lattice calculations, an approach which was
specifically designed for a systematic non-perturbative treatment of
QCD. In order to chart the progress made and to provide global
estimates for these quantities, the CKM-Lattice Working Group was
founded in February 2002 \cite{CKM02,lpl_lat02}. Here I report on
recent results and present global averages.

\section{Heavy quarks on the lattice} \label{sec_heavy}
%\vspace{-0.2cm}
Systematic effects in lattice simulations, as well as the more
specific problems of treating heavy quarks on the lattice have been
described many times in the literature (see
e.g. \cite{how97,joncts97}).

The great majority of lattice results for heavy-light decay constants
and $B$-parameters have to date been obtained in the quenched
approximation, where quark loops are neglected in the evaluation of
observables. The effects of non-zero lattice spacing $a$ (lattice
artefacts) have been studied extensively, though, and in many cases an
extrapolation to the continuum limit was performed. In order to
guarantee a smooth continuum behaviour, the non-perturbative
renormalisation of quark bilinears and four-fermion operators in the
discretised theory proved to be instrumental
\cite{MPSV94,alpha_lett95}. For some quantities, the level of
precision that can be reached in the continuum limit in the quenched
approximation is about 5\%. This then implies that current estimates
for decay constants are completely dominated by quenching effects. As
a consequence, most collaborations now focus on simulations with
$\nf=2$ or 3 dynamical quark flavours.

Another important systematic effect is the use of unphysical values of
the light quark masses $\mqu,\,\mqd$, both in quenched and unquenched
simulations. In particular, commonly used algorithms for dynamical
quarks slow down considerably for masses smaller than about
$\mqs/2$. Several proposals to address this problem have been made
\cite{qqq02,HasJan02,Daviesetal03,Luscher03}, but the effectiveness of
each approach must be studied in more detail. In order to make contact
with the chiral regime, one currently relies on extrapolations in the
light quark mass, using Chiral Perturbation Theory (ChPT) as a
guide. It has only been realised relatively recently that this can
introduce large uncertainties, since chiral logarithms are not
necessarily under control \cite{kr_02,panel_lat02,slov_02}.

Whenever one deals with heavy quarks on a lattice of spatial extent
$L$ and lattice spacing $a$, one is faced with a multi-scale problem,
in the sense that the following three inequalities cannot be satisfied
simultaneously:
\be
   am_{\rm b}\ll 1,\quad m_\pi{L}\gg 1,\quad L/a\;\lesssim\;50.
\ee
Violation of the first relation implies the presence of large lattice
artefacts, the second inequality must be satisfied if one aims for
small finite-volume effects, and the third is dictated by capacities
of current computers.

Several strategies to deal with this problem have been applied over
many years, among them the ``static approximation'' \cite{Eichten87},
the non-relativistic formulation (NRQCD) \cite{NRQCD}, the so-called
``Fermilab-approach'' \cite{EKM96}, and extrapolations in the heavy
quark mass.

\section{Recent results} \label{sec_results}
%\vspace{-0.2cm}
We now discuss results from several recent calculations, which will
also illustrate some of the issues raised earlier.

The first topic is a recent benchmark calculation of $\fDs$ in the
quenched approximation \cite{juettrolf03}. For lattice simulations,
the $D_{\rm s}$ meson is particularly appealing, since both the charm
and the strange quark can be treated directly, i.e. no extrapolations
are required to make contact with the physical values of the valence
quark masses. In ref. \cite{juettrolf03} the potentially large lattice
artefacts arising from relativistic $c$-quarks are eliminated through
an extrapolation to the continuum limit. By means of employing a
lattice action in which the leading lattice artefacts of O($a$) were
removed non-perturbatively \cite{imp:pap3}, the convergence of the
results to the continuum could be accelerated. Furthermore, the
authors used a non-perturbative estimate of the renormalisation factor
which connects the axial current on the lattice with its continuum
counterpart \cite{imp:pap4}.

The main result in ref. \cite{juettrolf03} is the continuum result for
$\fDs$ in units of the hadronic radius $r_0$ \cite{sommer_r093},
namely $r_0\fDs=0.638\pm0.024$, which translates into
\be
   \fDs=252\pm9\,\MeV
\ee
if the phenomenological value $r_0=0.5\,\fm$ is inserted. It is worth
emphasising that the only remaining uncertainty is due to quenching. A
crude estimate of the quenching error is obtained through the scale
uncertainty, i.e. the fact that different quantities yield different
estimates for the lattice scale in the quenched approximation. The
typical size of this ambiguity is about 10\% \cite{CPPACS_quen}.

In another recent calculation by de Divitiis \etal \cite{RomeII_FSS} a
new strategy to deal with the multi-scale problem was applied. Here
the condition $m_{\pi}L\gg 1$ was sacrificed in favour of $am_{\rm
b}\ll 1$. In this way one is able to accommodate a fully relativistic
$b$-quark, and a physically meaningful result is obtained if one
succeeds in determining the distortion due to the unphysically small
volume. The key observation is that this can be achieved through a
sequence of finite-size scaling steps, which relate the results
obtained for several lattice sizes $L_0,\,L_1,\ldots$:
\bea
   & &\fB(L_0)\;{\to}\;\fB(L_1)\;{\to}\;\fB(L_2)\;{\to}\;\ldots \\
   & &\fB(L_{k+1})=\fB(L_k)\,{\sigma_{\rm B}(L_k)},\quad L_{k+1}>L_k
\eea
Since the value of $\fB$ for the smallest volume, i.e. $\fB(L_0)$ and
the so-called ``step-scaling function'' $\sigma_{\rm B}(L_k)$ can be
computed in the continuum limit, one obtains a result for $\fB$ for
physically large volumes in a controlled manner. The main assumption,
which was verified explicitly in ref. \cite{RomeII_FSS}, is that the
finite-size effects depend only weakly on the heavy quark
mass. Starting from $L_0=0.4\,\fm$, a total of two scaling steps were
performed, each time doubling the lattice size. The initial value of
the decay constant in the continuum is $\fB(L_0)=471(2)\,\MeV$, and
together with the continuum estimates for the step scaling functions,
$\sigma_{\rm B}(L_0)=0.400(3)$ and $\sigma_{\rm B}(L_1)=0.92(4)$ one
obtains
\be
   \fB=173\pm8\pm4\,\MeV,
\ee
at $L_2=1.6\,\fm$, which is large enough to be identified with the
infinite volume limit. The quoted uncertainties cover all errors,
except that due to quenching. In a similar way the authors of
ref. \cite{RomeII_FSS} obtain
\bea
    & & \fBs=194\pm6\pm4\,\MeV,  \\
    & & \fD=217\pm7\pm5\,\MeV,\quad \fDs=239\pm5\pm5\,\MeV, \nonumber
\eea
the result for $\fDs$ being compatible with ref. \cite{juettrolf03}.

As mentioned earlier, the issue of chiral logarithms in
SU(3)-flavour-breaking ratios such as $\fBs/\fB$ or $\xi$ has
attracted a lot of attention recently. The dependence of heavy-light
decay constants on the mass of the light quark has usually been
modelled according to a naive linear ansatz, in which chiral
logarithms, as well as analytic terms arising at NLO in ChPT, are
neglected. The full expression at NLO reads
\bea
   & & {{\fBs\over\fB}}-1 = (\mK^2-\mpi^2){f_2(\mu)}  \nonumber \\
   & & -{{1+3{g^2}}\over{(4\pi f_\pi)^2}} \left[
         \frac{1}{2}I_{\rm P}(\mK)+\frac{1}{4}I_{\rm P}(m_\eta)
        -\frac{3}{4}I_{\rm P}(\mpi) \right]
\eea
where $I_{\rm P}(\mps)=\mps^2\ln(\mps^2/\mu^2)$ and $f_2$ is a
low-energy constant. As was pointed out by Kronfeld \& Ryan
\cite{kr_02}, the inclusion of chiral logarithms in the chiral
extrapolation of lattice data for heavy-light decay constants can
drastically change $\fBs/\fB$ and consequently also $\xi$, which
enters fits to the CKM parameters. By assuming
$g^2=g_{D^*D\pi}^2=0.35$ \cite{CLEO_gDDp} and $f_2=0.5(3)\,\GeV^{-2}$,
Kronfeld and Ryan conclude that $\xi=1.32\pm0.10$, which is more than
10\% larger than the global estimate quoted by Ryan in 2001
\cite{ryan_lat01}. It is somewhat ironic that the quantity $\xi$,
which for a long time had been assumed to be only weakly sensitive to
systematic effects, should be subjected to such a large
uncertainty. Note that the corresponding ratio of $B$-parameters,
$\BBs/\BB$ is largely unaffected, since the coefficient of the chiral
logarithm is $\propto(1-3g^2)$, which is close to zero.

\begin{figure}
\vspace{0.3cm}
%\resizebox{0.43\textwidth}{!}{ \includegraphics{Chi.phi_fB_b.eps} }
\psfig{file=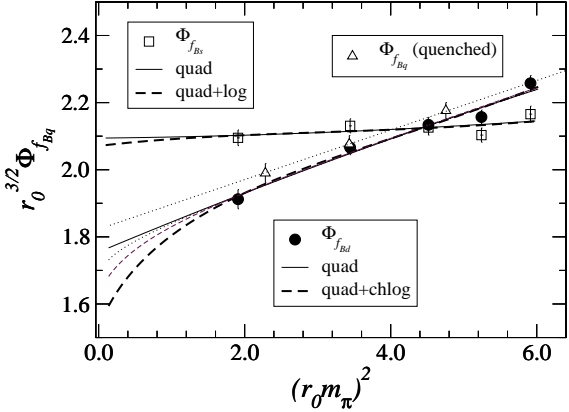,width=7.5cm}
%\special{psfile=./Chi.phi_fB_b.eps voffset -80 hoffset 0 vscale 43 hscale 43}
\caption{Dependence of $\Phi_{f_{\rm B}}$ and $\Phi_{f_{\rm{B_s}}}$ on
the light quark mass.}
\label{fig_JLQCD_BB}
\end{figure}

The predicted enhancement of $\fBs/\fB$ due to chiral logs seems
plausible, since the corresponding ratio in the light quark sector,
$\fK/f_\pi$, is known to come out too small in quenched QCD, if only a
naive linear quark mass dependence is assumed \cite{mbar:pap4}. In
ref. \cite{slov_02} it was argued that the chiral logarithms in
$\fBs/\fB$ and $\fK/f_\pi$ are nearly of the same size, so that one
would expect
%
%\be
   $\fBs/\fB\approx\fK/f_\pi=1.22$,
%\ee
%
which is compatible with \cite{kr_02}, but closer to previous global
estimates. The key question for any future determination is whether or
not the quark masses used in the simulation are light enough so that
ChPT at NLO gives a good description. In the context of $\fBs/\fB$
this was studied recently by the JLQCD Collaboration
\cite{JLQCD_BB_03}. The dependence of $\Phi_{f_{\rm B}}=f_{\rm
B}\sqrt{M_{\rm B}}$ and $\Phi_{f_{\rm B_s}}$ on the light quark mass
is shown in Fig. \ref{fig_JLQCD_BB}. JLQCD conclude that chiral
logarithms are not observed in the studied mass range. The modelling
of the effect due to chiral logs yields a drop of up to 11\% in $\fB$
relative to $\fBs$. The corresponding enhancement of $\fBs/\fB$ is of
the same order of magnitude as that of ref. \cite{kr_02}.

\section{Global estimates: Strategies \& Results} \label{sec_global}
%\vspace{-0.2cm}
Before I present global averages for decay constants and
$B$-parameters, I would like to outline the strategy which was used to
obtain them. The main point to note is that a global analysis of
lattice results is complicated by the fact that systematics can vary
substantially among results from different collaborations. The main
differences lie in the treatment of lattice artefacts, the choice of
quantity that sets the lattice scale, the renormalisation procedure
for currents and four-fermion operators on the lattice, and the
details of extrapolations in the quark masses, either down to the
chiral regime, or, where applicable, to the mass of the
$b$-quark. Therefore, a straightforward global average of results is
unreliable.

For the procedure applied here I have decided to focus on the quenched
approximation: although unphysical, the quenched approximation is
known to describe the light hadron spectrum at the level of 10\%
\cite{CPPACS_quen}. Furthermore, a wealth of results is available for
$\nf=0$: the continuum limit has been taken in almost all recent
quenched calculations, and other systematic effects have been studied
thoroughly.

Simulations with dynamical quarks, either for $\nf=2$ or~3 have not
yet reached the same level of maturity. The masses of the sea quarks
are still quite large, so that their effects on observables are likely
to be suppressed. Performing reliable continuum extrapolations is much
more expensive, all of which implies that a clear separation of
sea quark effects from lattice artefacts is very difficult. There are,
however, attempts to expose the effects of unquenching by focusing on
ratios in which systematic uncertainties other than quenching largely
cancel. For instance, the CP-PACS \cite{CPPACS_FNAL_00} and MILC
\cite{MILC_fB_PRD} collaborations have found
%
%\be
   ${\fDs^{\nf=2}}/{\fDs^{\nf=0}}\approx 1.10$ at $a\approx0.1\,\fm$,
%\label{eq_fDs_ratio}
%\ee
%
which implies a 10\%-enhancement in $\fDs$ due to dynamical quark
effects. No extrapolations in the valence quark masses are required,
but it remains to be seen whether this number is stable against
lattice artefacts and whether lighter sea quarks lead to a significant
change.

As regards chiral logarithms in ratios like $\fBs/\fB$ and $\xi$,
their effects have not been quantified from first principles so far. In
what follows, I shall therefore use results assuming the naive (LO)
dependence on the light quark mass, but allow for a 10\% {\it
increase} in $\fBs/\fB$ and $\fDs/\fD$.

\begin{figure*}
\leavevmode
\psfig{file=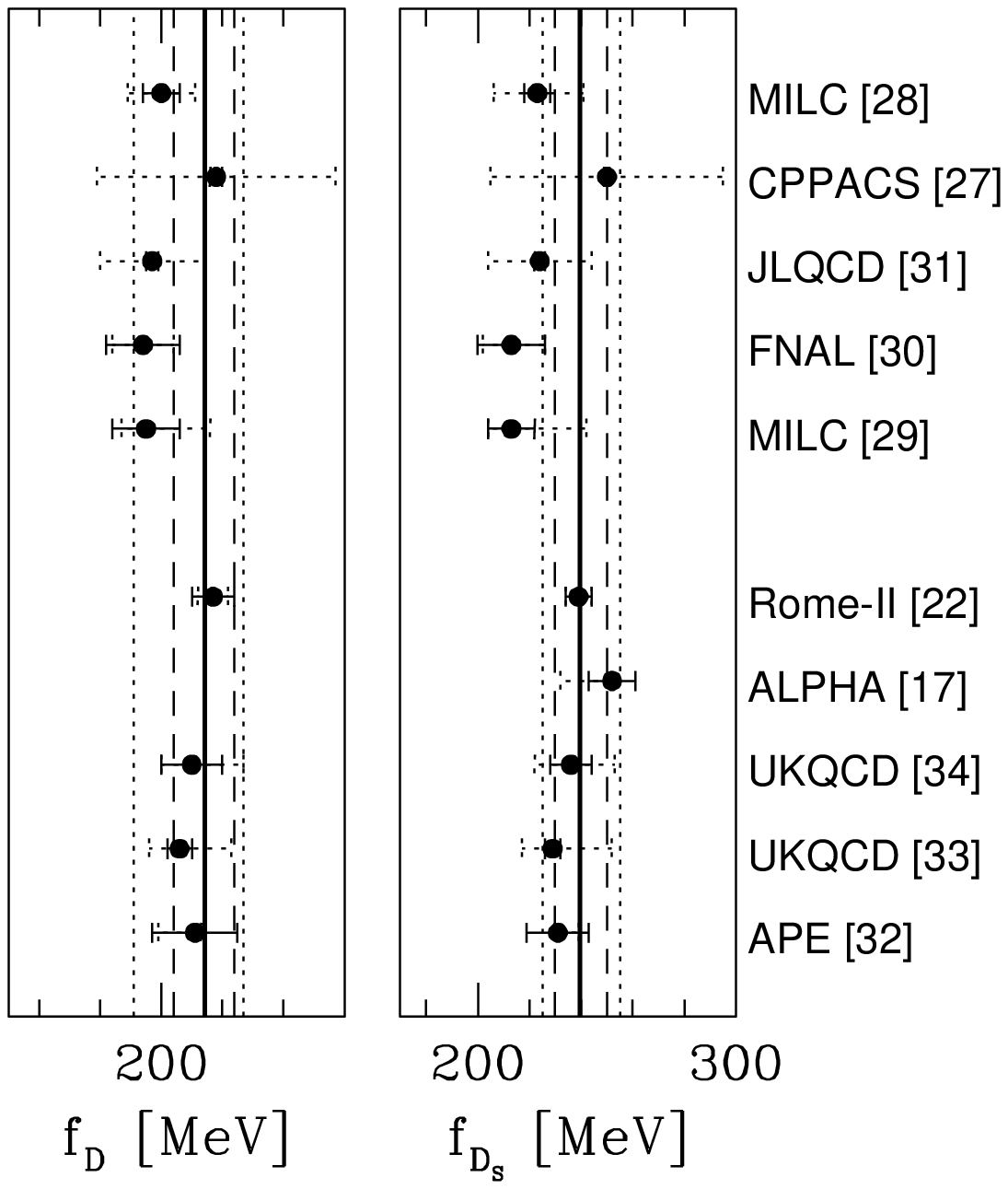,width=5.8cm}
\psfig{file=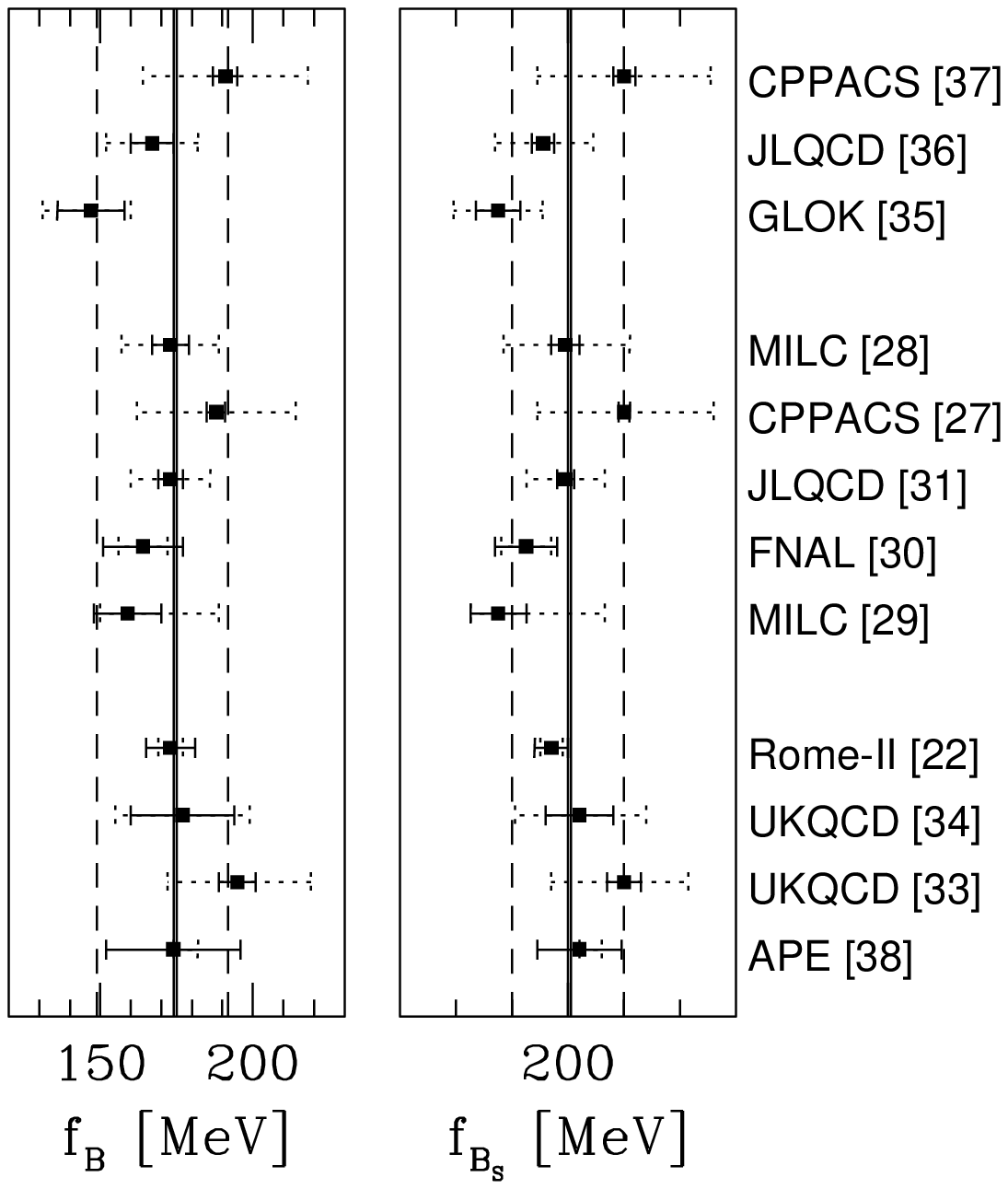,width=5.8cm}
\psfig{file=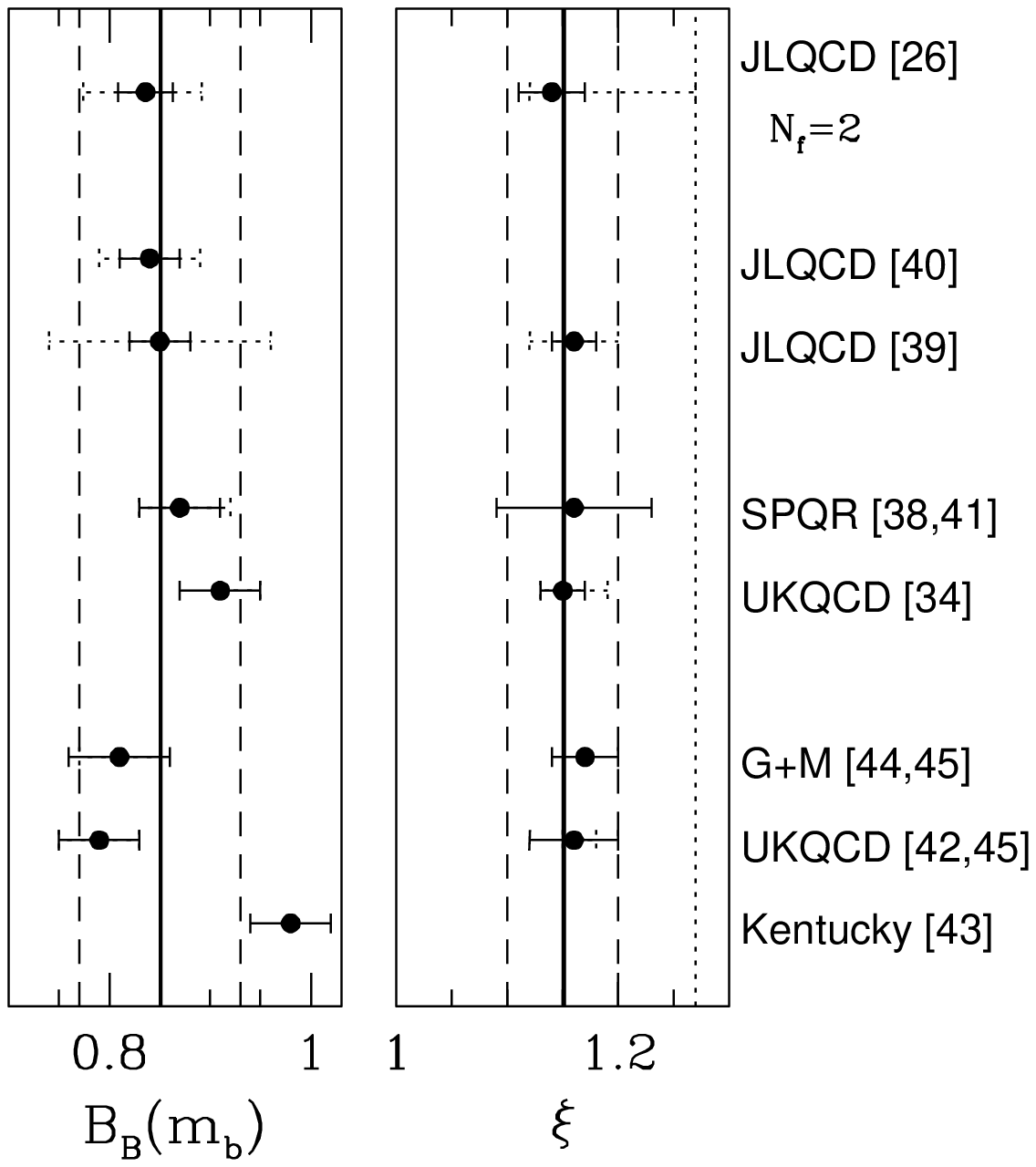,width=5.8cm}
\caption{Recent results for decay constants and $B$-parameters in the
quenched approximation, compared to the global averages of
eqs.(\ref{eq_fds_quen}), (\ref{eq_fdd_quen}), (\ref{eq_fbs_quen}),
(\ref{eq_fbd_quen}) and (\ref{eq_BB_rep}) (solid vertical lines). For
$D$-mesons the dashed lines indicate the statistical error, while the
systematic error is represented by the dotted lines. For the
$B$-system the dashed vertical lines denote the total error, with the
uncertainty due to chiral logs included in quadrature. Dotted error
bars on individual determinations indicate the quoted systematic
error. The result for $\BB$ and $\xi$ in \cite{JLQCD_BB_03} was
obtained for $\nf=2$.}
\label{fig_fds_quen}
\end{figure*}

The starting point for the global analysis is $\fDs$ in the quenched
approximation ($\nf=0$). The recent benchmark calculation
\cite{juettrolf03} demonstrates that the continuum value in quenched
QCD (for a given quantity that sets the scale) is obtained with a
total accuracy of 4\%. The central value may vary between 225 and
255\,\MeV, depending on the chosen scale. Thus we have
\be
   \fDs^{\nf=0}=240\pm10\,\hbox{(stat)}\pm15\,\hbox{(scale)}\,\MeV.
\label{eq_fds_quen}
\ee
In order to estimate $\fD$ one can divide this result by
$\fDs/\fD=1.12\pm0.02{}^{+0.11}_{-0.00}$. This is the number quoted by
Ryan \cite{ryan_lat01} for $\nf=0$, except for the additional 10\%
asymmetric uncertainty due to chiral logs. This yields
\be
   \fD^{\nf=0}=214\pm10\,\hbox{(stat)}\pm13\,\hbox{(scale)}{}^{+~0}_{-19}
                        \,\hbox{($\chi$log)}\,\MeV.
\label{eq_fdd_quen}
\ee
In Fig.\,\ref{fig_fds_quen} the quenched global estimates of
eqs.\,(\ref{eq_fds_quen}) and~(\ref{eq_fdd_quen}) are compared with
recent determinations from various groups, who have used either the
Fermilab approach
\cite{MILC_fB_PRL,FNAL_98,JLQCD_98,CPPACS_FNAL_00,MILC_fB_PRD}, or the
O($a$) improved Wilson action
\cite{APE_99,UKQCD_01,UKQCD_LL01,juettrolf03,RomeII_FSS} to treat the
heavy quark. Note that the results from different collaborations have
not been converted to a common lattice scale. Without this conversion
it is not clear whether the observed spread is just indicative of the
scale ambiguity, or due to some other, possibly uncontrolled
systematic effect.

The next step is to account for dynamical quarks. In my view the
safest procedure is to multiply the quenched values of $\fD$ and
$\fDs$ by the ratio
%(c.f. eq.~(\ref{eq_fDs_ratio}))
%
\be
   f_{\rm P_q}^{\nf=2}/f_{\rm P_q}^{\nf=0} = 1.10\pm0.05,
   \quad P=D,B,\quad q=d,s,
\label{eq_nf2and0}
\ee
in which some of the systematic errors can be expected to cancel. The
number in eq.~(\ref{eq_nf2and0}) is based on observations in
ref. \cite{CPPACS_FNAL_00} that dynamical quarks enhance the values of
decay constants by $5-15$\,\%, largely independent of the valence
quark contents. Thus we find
\bea
 & &\fDs^{\nf=2}=264\pm11\,\hbox{(stat)}\pm22\,\hbox{(quen)}\,\MeV \\
 & &f_{\rm D}^{\nf=2}=235\pm11\,\hbox{(stat)}\pm19\,\hbox{(quen)}\,
                        {}^{+~0}_{-21}\,\hbox{($\chi$log)}\,\MeV,\nonumber
\eea
where I have combined the scale uncertainty and the error in
eq. (\ref{eq_nf2and0}) into a total quenching error. Here we make no
attempt to try and estimate the effects of a dynamical strange quark,
and hence the above numbers are our final results for $\fD$ and $\fDs$
which we list once more in Table~\ref{tab_global}.

The procedure to obtain estimates for $\fB$ and $\fBs$ is entirely
analogous, with one important difference: unlike the case of
$D$-mesons, there is not yet a quenched benchmark calculation of
$\fBs$, which would yield the total error within the quenched
approximation. As a starting point we therefore use the ``global
representation'' of quenched results quoted by Ryan \cite{ryan_lat01},
i.e.
\be
   \fBs^{\nf=0}=200\pm20\,\MeV.
\label{eq_fbs_quen}
\ee
Here the error has been obtained by requiring consistency with a
number of different results being subjected to different systematics.
Dividing by \cite{ryan_lat01}
\be
    \fBs/\fB=1.15\pm0.03{}^{+0.12}_{-0.00}\,\hbox{($\chi$log)}
\label{eq_fBsoverfB}
\ee
gives
\be
   \fB^{\nf=0}=174\pm18\,{}^{+~0}_{-17}\,\hbox{($\chi$log)}\,\MeV,
\label{eq_fbd_quen}
\ee
and multiplication with the ratio of eq. (\ref{eq_nf2and0}) yields the
numbers listed in Table \ref{tab_global}. In Fig. \ref{fig_fds_quen}
the global quenched estimates are again compared to individual results
obtained using NRQCD \cite{GLOK_SGO_98,JLQCD_fB_00,CPPACS_NRQCD_01},
the Fermilab approach
\cite{MILC_fB_PRL,FNAL_98,JLQCD_98,CPPACS_FNAL_00,MILC_fB_PRD}, the
finite-size scaling technique \cite{RomeII_FSS}, as well as
extrapolations in the heavy quark mass from the region of the charm
quark mass \cite{APE_01,UKQCD_01,UKQCD_LL01}.

\begin{table}
\caption{Global estimates for decay constants and $B$-parameters. The
numbers are understood to refer to full QCD.}
\label{tab_global}
\begin{tabular}{l}
\hline\noalign{\smallskip}
$\fDs=264\pm11\pm22\,\hbox{(quen)}\,\MeV$ \\
$f_{\rm D_{\phantom{\rm s}}}=235\pm11\pm19\,\hbox{(quen)}\,
                        {}^{+~0}_{-21}\,\hbox{($\chi$log)}\,\MeV$ \\
$\fBs=220\pm25\,\MeV$ \\
$f_{\rm B_{\phantom{\rm s}}}=191\pm23\,
                        {}^{+~0}_{-19}\,\hbox{($\chi$log)}\,\MeV$ \\ \\
$\BBhat=\BBshat=1.34\pm0.12$ \\
$\xi=1.15\pm0.05\,{}^{+0.12}_{-0.00}$ \\
$\fBs\sqrt{\BBshat}=255\pm31\,\MeV$ \\
$f_{\rm B_{\phantom{\rm s}}}\sqrt{\widehat{B}_{\rm B_{\phantom{\rm s}}}}
           =221\pm28\,{}^{+~0}_{-22}\,\hbox{($\chi$log)}\,\MeV$ \\
\noalign{\smallskip}\hline
\end{tabular}
\end{table}

For the $B$-parameters $\BB$, $\BBs$ and the ratio $\xi$, there are
not so many results available, and only one recent calculation uses
dynamical quarks \cite{JLQCD_BB_03}. It is then clear that systematics
cannot be studied as thoroughly as for decay constants. In particular,
the continuum limit has not been taken in any study so far. It turns
out, though, that results for $B$-parameters and $\xi$ are broadly
consistent, regardless of whether NRQCD
\cite{JLQCD_BB_00,JLQCD_BB_02,JLQCD_BB_03}, relativistic heavy quarks
\cite{UKQCD_LL01,APE_01,SPQR_02} or the static approximation
\cite{UKQCD_95,Kent_97,GandM_97,GandR_99} are used in
simulations. Apparently, dynamical quarks do not lead to a significant
enhancement of $\BB$ or $\BBs$ \cite{JLQCD_BB_03}, contrary to what is
observed for decay constants. Chiral logarithms in the SU(3)-flavour
breaking ratio $\BBs/\BB$ are suppressed, and a linear extrapolation
in the light quark mass yields a value compatible with one. All
published data are then consistent with the global representation:
\bea
%  & & \BB(m_b)=0.85\pm0.08\;\Rightarrow\;\BBhat^{\rm NLO}= 
%      1.34\pm0.12  \label{eq_BB_rep} \\
%  & & \BBs/\BB=1.00\pm0.03,\quad \xi=1.15\pm0.05{}^{+0.12}_{-0.00}\,
%      \hbox{($\chi$log)}. \nonumber
  & & \BB(m_b)\;=0.85\pm0.08\;\Rightarrow\;\BBhat^{\rm NLO}= 
      1.34\pm0.12  \nonumber \\
  & & \BBs(m_b)=0.85\pm0.08\;\Rightarrow\;\BBshat^{\rm NLO}= 
      1.34\pm0.12  \label{eq_BB_rep} \\
  & & \BBs/\BB=1.00\pm0.03,\quad \xi=1.15\pm0.05{}^{+0.12}_{-0.00}\,
      \hbox{($\chi$log)}. \nonumber
\eea
The value of $\xi$ is obtained by combining eq. (\ref{eq_fBsoverfB})
with the above result for $\BBs/\BB$. The collection of global
estimates in Table \ref{tab_global} is consistent with refs.
\cite{CKM02,ryan_lat01,becir_ckm03}.

\section{Future studies}
\vspace{-0.1cm}
How can the current global estimates be improved in order to sharpen
the constraints on CKM parameters? There are several areas in which
progress can be made. 

Future efforts must surely focus on dynamical simulations, but the
quenched approximation remains helpful for the understanding of
several issues. In my view, a quenched benchmark value of $\fBs$ is
very important to quantify the uncertainties arising from
discretisation and renormalisation effects. Although this is precisely
the aim of the finite-size scaling method of \cite{RomeII_FSS}, an
independent check using a different method should be performed. Here
one may think of an interpolation between results obtained in the
static approximation and those obtained near $m_{\rm c}$, {\it after}
the continuum limit has been taken. With the advent of methods that
allow for precise non-perturbative determinations of the
renormalisation factor for the axial current \cite{ALPHA_ZAstat}, as
well as for a better signal/noise ratio in the static approximation
\cite{ALPHA_SigNoi}, one can obtain results with much higher accuracy
than previously possible \cite{Heitger_eps03}. This strategy may also
be extended to $B$-parameters \cite{ALPHA_prep}.

Obviously the most pressing problem is to simulate small dynamical
quark masses more efficiently. The r\^ole of ``improved staggered
quarks'' has been vigorously emphasised \cite{Daviesetal03} in this
context, with first results being reviewed in
\cite{sg_lat03}. Eventually such efforts should yield not only
estimates for, say, $\fBs^{\nf=3}$, but should also settle the issue
of chiral logarithms in $\fBs/\fB$, $\xi$ and $\fK/f_\pi$. The latter
is, in fact, explored separately by a number of groups using different
algorithms and discretisations \cite{qqq03,JLQCD_BB_03}. In this
context, the so-called Grinstein ratio
%
%\be
   $\left({\fBs/\fB}\right)\left/\left({\fDs/\fD}\right)\right.$
%\ee
%
in which the chiral logarithms cancel \cite{Grin94}, has been
suggested as a more reliable way to determine $\fBs/\fB$ on the
lattice \cite{JLQCD_BB_03}, once $\fDs$ and $\fD$ have been measured
at high-luminosity charm factories such as CESR-c \cite{ckm_wg5}. This
approach, however, can only succeed if the quark masses used in
simulations are light enough so that the data are consistently
described by ChPT at NLO.

Finally, one may think of better ways to obtain global averages. This
may entail potentially laborious re-analyses of existing simulation
data. In order to facilitate a comparison all results should be
converted to a common scale, such as $r_0$. Systematic errors should
best be quantified after an extrapolation to the continuum limit. Here
on might concentrate on ratios like $\fBs/\fDs$ for fixed $\nf$, or
$\fDs^{\nf=2}/\fDs^{\nf=0}$. Thereby one has a better chance to expose
the effects of replacing $c$ by $b$-quarks and those due to dynamical
quarks.
\vspace{0.3cm}

It is a pleasure to thank P. Hern\'andez, M. Papinutto and C. Pena for
helpful discussions. I thank the organisers as well as the conveners
V. Lubicz and B. Pioline for a stimulating conference. I am grateful
to I. Montvay and C. Pena for a critical reading of the manuscript.

\end{document}